\documentclass[conference]{IEEEtran}

\IEEEoverridecommandlockouts

\IEEEpubid{\makebox[\columnwidth]{979-8-3503-7550-3/24/\$31.00 \copyright 2024 IEEE \hfill}
\hspace{\columnsep}\makebox[\columnwidth]{ }}
\usepackage{cite}
\usepackage{svg}
\usepackage{amsmath,amssymb,amsfonts}
\usepackage{graphicx}
\usepackage{textcomp}
\usepackage{xcolor}
\usepackage{tabularx}
\usepackage{makecell}
\def\BibTeX{{\rm B\kern-.05em{\sc i\kern-.025em b}\kern-.08em
  T\kern-.1667em\lower.7ex\hbox{E}\kern-.125emX}}
\usepackage{booktabs}
\usepackage{algpseudocode}
\usepackage{amsmath}
\usepackage{amssymb}
\usepackage{multirow}
\usepackage{subcaption}
\usepackage{hyperref}
\usepackage[numbers, sort&compress]{natbib}
\begin{document}

\title{\textcolor{black}{AXAI-CDSS : An Affective Explainable AI-Driven Clinical Decision Support System for Cannabis Use}}
\author{\IEEEauthorblockN{Tongze Zhang \href{https://orcid.org/0000-0002-3375-7136}{\includegraphics[scale=0.06]{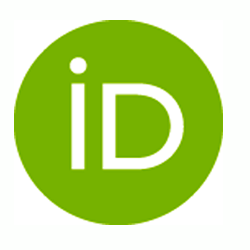}}}
\IEEEauthorblockA{
\textit{Stevens Institute of Technology}\\
Hoboken, New Jersey }
\and

\IEEEauthorblockN{Tammy Chung \href{https://orcid.org/0000-0002-1527-2792}{\includegraphics[scale=0.06]{graph/orcid.png}}}
\IEEEauthorblockA{
\textit{Rutgers University}\\
Newark, New Jersey }
\and
\IEEEauthorblockN{Anind Dey \href{https://orcid.org/0000-0002-3004-0770}{\includegraphics[scale=0.06]{graph/orcid.png}}}
\IEEEauthorblockA{
\textit{University of Washington}\\
Seattle, Washington}
\and
\IEEEauthorblockN{*Sang Won Bae \href{https://orcid.org/0000-0002-2047-1358}{\includegraphics[scale=0.06]{graph/orcid.png}}}
\IEEEauthorblockA{
\textit{Stevens Institute of Technology}\\
Hoboken, New Jersey}
}
\maketitle

\begin{abstract}
\textcolor{black}{
As cannabis use has increased in recent years, researchers have come to rely on sophisticated machine learning models to predict cannabis use behavior and its impact on health. However, many artificial intelligence (AI) models lack transparency and interpretability due to their opaque nature, limiting their trust and adoption in real-world medical applications, such as clinical decision support systems (CDSS). To address this issue, this paper enhances algorithm explainability underlying CDSS by integrating multiple Explainable Artificial Intelligence (XAI) methods and applying causal inference techniques to clarify the models’ predictive decisions under various scenarios. By providing deeper interpretability of the XAI outputs using Large Language Models (LLMs), we provide users with more personalized and accessible insights to overcome the challenges posed by AI's “black box” nature. Our system dynamically adjusts feedback based on user queries and emotional states, combining text-based sentiment analysis with real-time facial emotion recognition to ensure responses are empathetic, context-adaptive, and user-centered. \textcolor{black}{This approach bridges the gap between the learning demands of interpretability  and the need for intuitive understanding, enabling non-technical users such as clinicians and clinical researchers to interact effectively with AI models.} Ultimately, this approach improves usability, enhances perceived trustworthiness, and increases the impact of CDSS in healthcare applications.}

\end{abstract}

\begin{IEEEkeywords}
Explainable Artificial Intelligence (XAI), Passive Sensing, Affective Computing, Clinical Decision Support Systems (CDSS), Cannabis Use Disorder, Cannabis Intoxication, Cannabis-Intoxicated Behaviors, Personalized Intervention, Large Language Models (LLMs), Algorithmic Decisions, Transparency, Healthcare AI, Trustworthy AI, Facial Emotion Recognition, Causal Inference
\end{IEEEkeywords}

\section{Introduction} 
The prevalence of cannabis use has risen dramatically over the past decade, leading to an increasing number of people meeting criteria for Cannabis Use Disorder (CUD) \citep{elhaddad2024ai,olyanasab2024leveraging}. This surge presents significant challenges for healthcare providers, who must develop accurate and reliable methods for diagnosis and treatment \cite{golden2024applying}. To address these challenges, Artificial Intelligence (AI) and Machine Learning (ML) technologies have emerged as promising tools in healthcare, offering new ways to analyze complex data and providing insights that may not be readily apparent through traditional methods \cite{lourdusamy2020clinical}. AI has great potential to revolutionize clinical decision-making; however, the “black box" nature of many AI models hinders their application in clinical settings. This opacity in the AI decision-making process often leads to skepticism and mistrust among healthcare providers, who may be reluctant to rely on these systems to make critical clinical decisions \cite{golden2024applying}.



Traditionally, drug use prediction models in addiction research have relied on isolated data sources, either focusing on verbal cues, self-reports, or independently on sensor signals \cite{lourdusamy2020clinical}. However, the complexity of substance use behaviors requires a more holistic, multimodal approach to capturing patterns of substance use in an individual’s daily life \cite{elhaddad2024ai} . Motivated by \cite{10652070} \cite{bae2025enhancing}, we combine linguistic data collected through user interactions on a customized platform with continuous sensor data from smartphones and wearable devices, aiming to create an integrated system capable of real-time prediction and intervention.


In order to reduce skepticism due to the opacity of AI models, the concept of explainable AI (XAI) has received attention as it aims to make AI systems more transparent and understandable to end users \cite{arrieta2020explainable}. This transparency enhances trust and facilitates the integration of AI into clinical practice. The aim of XAI is to make AI systems more transparent and explainable \cite{arrieta2020explainable}, thereby enhancing trust and facilitating their integration into clinical practice. These insights enable clinicians to understand the reasoning behind AI recommendations, thereby increasing confidence and trust \cite{tonekaboni2019clinicians} in using these systems as part of the decision-making process. Additionally, integrating sentiment analysis into clinical decision support systems (CDSS) is a new way to enhance the interaction between healthcare providers and AI systems \cite{denecke2015sentiment}. Sentiment analysis enables the system to determine the user's emotional state and thus respond accordingly. This personalized interaction increases user engagement \cite{bickmore2005establishing}, by creating a more empathetic and supportive clinical environment \cite{del2012relationship}.



The novelty of this study lies in the joint use of multiple models. The first model uses facial expression recognition technology to analyze the emotional expression of the person (e.g., healthcare provider) using the system, and text sentiment analysis to identify the emotional state of the user. A second large language model focuses on generating interpretable AI-based explanations and clinical recommendations. This innovative combination of emotional state analysis and generation of AI-based explanation and recommendations not only ensures clear delineation of tasks, but also improves the overall efficiency and effectiveness of the system. By integrating emotion recognition of the user (i.e., the clinician) into the CDSS, the system dynamically adapts tone and content to the clinician's emotional state, providing empathic responses that promote trust and reduce cognitive load in busy clinical environments. At the same time, the interpretable AI component ensures transparency in the decision-making process, addressing the long-standing “black box” problem of AI-based CDSS.

The main objective of this research is to construct an affective-adaptive CDSS to help users (i.e., clinicians) obtain more effective AI interpretive feedback. While the system is designed to perform well in high-stress situations commonly encountered in clinical settings, it is also tailored to adapt to varying levels of a user's emotional states, ensuring usability and effectiveness even when clinicians are not operating under high stress. This approach accounts for the potential variability in affect when clinicians interact with the system \cite{elhaddad2024ai}. We integrate bilingual models for affective analysis and interpretive feedback generation, respectively, and provide personalized affective-adaptive responses in user interactions \cite{umerenkov2023deciphering}. This research explores how to optimize the user experience through human-computer interaction by developing an application system that integrates interpretive AI and affective-driven feedback  \cite{elhaddad2024ai}.
In this study, we will discuss the design of the user interaction platform and address ethical issues regarding data privacy and participant consent. Our findings contribute to the growing body of research on substance use intervention and highlight the potential of combining advanced AI techniques with mobile health and data from wearables to support clinical interventions. \textcolor{black}{The system is intended to assist clinicians in decision-making, rather than requiring them to make diagnoses based on emotional states. Its main purpose is to provide explainable insights and emotional adaptation based on artificial intelligence to support clinicians working in high-stress environments.}

This study answers several key research questions:
How does the integration of two language models and sentiment analysis impact the user experience and clarity of AI explanations in CDSS?



\textcolor{black}{
What are the key challenges and benefits of integrating real-time emotion recognition to improve user interaction and decision-making in AI-driven clinical systems?}
\textcolor{black}{
How does the variability in clinicians' emotional responses, particularly for those new to the system, affect their interpretation and adoption of AI-driven insights?}





\section{Related Work}

\subsection{Advantages of Language Models in designing CDSS-HCI}

Clinical Decision Support Systems (CDSS) developed using Artificial Intelligence (AI) and Machine Learning (ML) technologies have received much attention in recent years, especially in improving healthcare outcomes \cite{rajashekar2024human}. The integration of AI into healthcare systems is driven by the need to improve the accuracy, efficiency, and personalization of patient care. However, the adoption of AI in clinical settings has been challenging, especially due to the opaque nature of many AI models, leading to concerns about their trustworthiness and interpretability. This section reviews related work in the field of explainable AI (XAI) \cite{antoniadi2021current} and sentiment analysis \cite{grasser2018aspect} and its application in healthcare, particularly in the context of cannabis use disorder (CUD) \citep{mennis2023cannabis, ding2024spatial}.

Language models have recently gained prominence as a powerful tool for designing CDSS within the Human-Computer Interaction (HCI) framework. The main advantages of Language Models include the ability to process and analyze large amounts of unstructured textual data to provide valuable insights and predictions. Approaches often include training these models on large datasets to support clinical decision making \cite{dayanandan2024enabling}. The results of several studies have shown that language models can significantly improve diagnostic accuracy, personalize patient care, and provide decision support by identifying patterns and correlations in data that may have been missed by clinicians \citep{rajashekar2024human, tazin2024understanding}. However, language models also have significant limitations, including the need for large amounts of high-quality training data, the possibility of inherent bias in models, and challenges related to the interpretability and transparency of the decision-making process. Addressing these limitations is critical to the reliable and ethical use of language models in clinical settings  \citep{wilhelm2023large,luo2024clinical}.

This predictive capability is particularly beneficial for chronic disease management, as early detection and timely intervention are critical to preventing disease progression and improving patient prognosis.


\subsection{XAI in CDSS-HCI}
Explainable Artificial Intelligence (XAI) has become an important area of research that addresses the "black box" problem associated with many machine learning models \cite{arrieta2020explainable}. The need for transparency in the AI decision-making process is particularly acute in healthcare, where clinicians need a clear understanding of the factors that drive AI-generated recommendations. Various XAI techniques have been proposed to improve the interpretability of AI models and provide clinicians with the insights needed to trust and effectively use these systems. In this study, although we explored a variety of XAI techniques including SHAP (Shapley Additive Explanations) \cite{lundberg2020local}, rule-based explanations, and counterfactual explanations \cite{wachter2017counterfactual}, we also use causal learning models \cite{zheng2024causal} to improve our analyses and interpretation of the data.

\subsection{Prompt Engineering for AI Interactions}
In recent years, in the field of Natural Language Processing (NLP), Prompt Engineering has emerged as a key technique for guiding language models to generate specific outputs. Some research has been done in few-shot learning and zero-shot learning modes, where the user can provide a small number of textual hints to guide the model in generating a solution to a complex task. Compared to traditional fine-tuning methods, prompt engineering reduces time and cost, allowing the model to adapt to different task requirements without further training \cite{brown2020language}. These studies show that with well-designed cues, language models can be guided to perform multiple tasks.

\textcolor{black}{As dialog systems are more and more widely used in HCI scenarios, users expect AI-driven systems to not only generate content-correct responses, but also to understand and adapt to their emotional states. Existing research emphasizes that emotional states such as anger, disgust, fear, happiness, sadness, surprise, and neutrality are typically expressed during human-computer interactions. Systems that are able to reliably detect and respond to these emotions show the potential to improve user satisfaction by increasing the perceived relevance and empathy of their responses.} Therefore, prompt engineering has been gradually used as an effective means to build emotionally adaptive dialog systems \cite{li2024cfn}. This emotion-adaptive prompt design not only enhances the naturalness of the dialog, but also improves user satisfaction in emotional interactions \cite{wen2024personality}.

\subsection{Facial emotion recognition in CDSS}

\textcolor{black}{Integrating facial emotion recognition into CDSS has emerged as a promising advancement for enhancing the interaction between healthcare providers and AI tools. Facial emotion recognition utilizes computer vision and machine learning techniques to detect and interpret human facial expressions, enabling systems to assess a user's emotional state in real time \cite{liu2024affective}. This capability is particularly valuable in high-stress healthcare environments, where clinicians often experience a range of emotions that can affect their decision-making processes. While this ability may be valuable in medical settings where clinicians may face high workloads or manage critical patient cases, it can also be adapted to different levels of stress and emotional states. The use of facial emotion recognition in CDSS is in line with the principles of human-computer interaction (HCI) and affective design, which emphasize the importance of user-centered and emotionally intelligent interfaces \cite{huang2023emotion}. This approach improves system usability and promotes a stronger trusting relationship between clinicians and AI tools. By recognizing and responding to users' emotional states, CDSS can provide a more personalized and effective support experience \cite{peter2008affect}.} 

Many systems focus solely on providing evidence-based recommendations without considering the emotional state of the clinician, \textcolor{black}{which can have an impact on the decision-making process, both in high-stress environments and in routine clinical settings. The system is designed to support clinicians by adapting to different levels of user stress, ensuring that the clinician is able to make optimal evidence-based recommendations in a variety of situations \cite{pepa2021automatic}.} Although some studies have incorporated text sentiment analysis to adapt system responses, these approaches often neglect non-verbal cues, which are critical to a full understanding of a user's emotions \cite{pereira2024systematic}. In addition, existing emotion recognition systems often rely on a single modality—text or facial expressions—which may not capture the full range of emotional cues. This single modality approach may result in an incomplete or inaccurate assessment of a user's emotional state \cite{ganapathy2021emotion}.

\subsection{Gaps in health risk monitoring, prediction and intervention}

Despite significant progress, large gaps remain in health risk surveillance, prediction and intervention. Current systems often face challenges in integrating with various data sources, maintaining data accuracy and providing real-time updates. Many existing health monitoring systems are heavy on analyzing data from a single data source such as electronic health records (EHRs), wearable devices, or other health-related technologies. This fragmentation limits the effectiveness of predictive models and the ability to provide real-time health updates \citep{wang2023explainable, liao2024ehr}. Current research on addiction exhibits several gaps, particularly in the integration of natural language processing (NLP) and machine learning (ML) to fully understand and predict addiction-related health risks. Several studies have shown the potential of NLP in processing clinical data to predict addiction relapse and other health outcomes (e.g., \cite{kramer2024analysis}). Another study highlighted the role of ML in analyzing social media data to detect patterns related to substance use and mental health \cite{chhetri2023machine}. However, the integration of these approaches remains underexplored.

Although the potential of combining NLP and ML to enhance addiction research is well recognized, research that fully integrates these techniques remains scarce. 
Integrating NLP and ML in a clinical decision support system is critical. Machine learning models, combined with XAI and causal modeling, provide transparent, interpretable analytics for patient data, enabling healthcare providers to understand complex algorithms, thereby enhancing the decision-making process. Further the use of language models allows systems to understand and generate natural language, making it easier for clinicians to interpret results. Most of the existing research tends to focus on NLP or ML individually, rather than exploring the synergistic effects of using them together \citep{henry2021natural,harrison2021machine}. The lack of an integrated approach has resulted in a lack of fully utilizing these technologies to better understand, predict, and intervene in addiction-related health risks \cite{sutton2020overview}. Sentiment analysis is a tool in the field of NLP traditionally used to assess the sentiment tone of textual data. However, the use of sentiment analysis in healthcare, particularly in CDSS, introduces a new dimension to the interaction between patients and clinicians \cite{Denecke2023}. In contrast to the traditional use of sentiment analysis for understanding patient emotions, in our CDSS, sentiment analysis aims to match and respond to the emotional state of clinicians. The reason for applying sentiment analysis to clinicians is that the emotional and psychological state of healthcare providers can greatly influence their decision-making process and overall effectiveness \cite{schutze2023requirements}. High-stress environments, emotional exhaustion, and the complexity of clinical decision-making all affect how clinicians interact with AI systems. Therefore, a sentiment analysis module that detects and aligns with a clinician's emotional state plays a critical role in improving the usability and acceptance of a CDSS \cite{sutton2020overview}. This alignment can make the interaction more enjoyable for the clinician, and also help to reduce cognitive load, thus allowing the clinician to more effectively  focus on patient care. Additionally, by providing emotionally intelligent feedback, the system fosters a collaborative relationship between the clinician and the AI, reducing user frustration and thus better integrating AI recommendations into the clinical workflow \cite{schutze2023requirements}. 
\section{Method}

\begin{figure*}
  \centering
  \includegraphics[width=\linewidth]{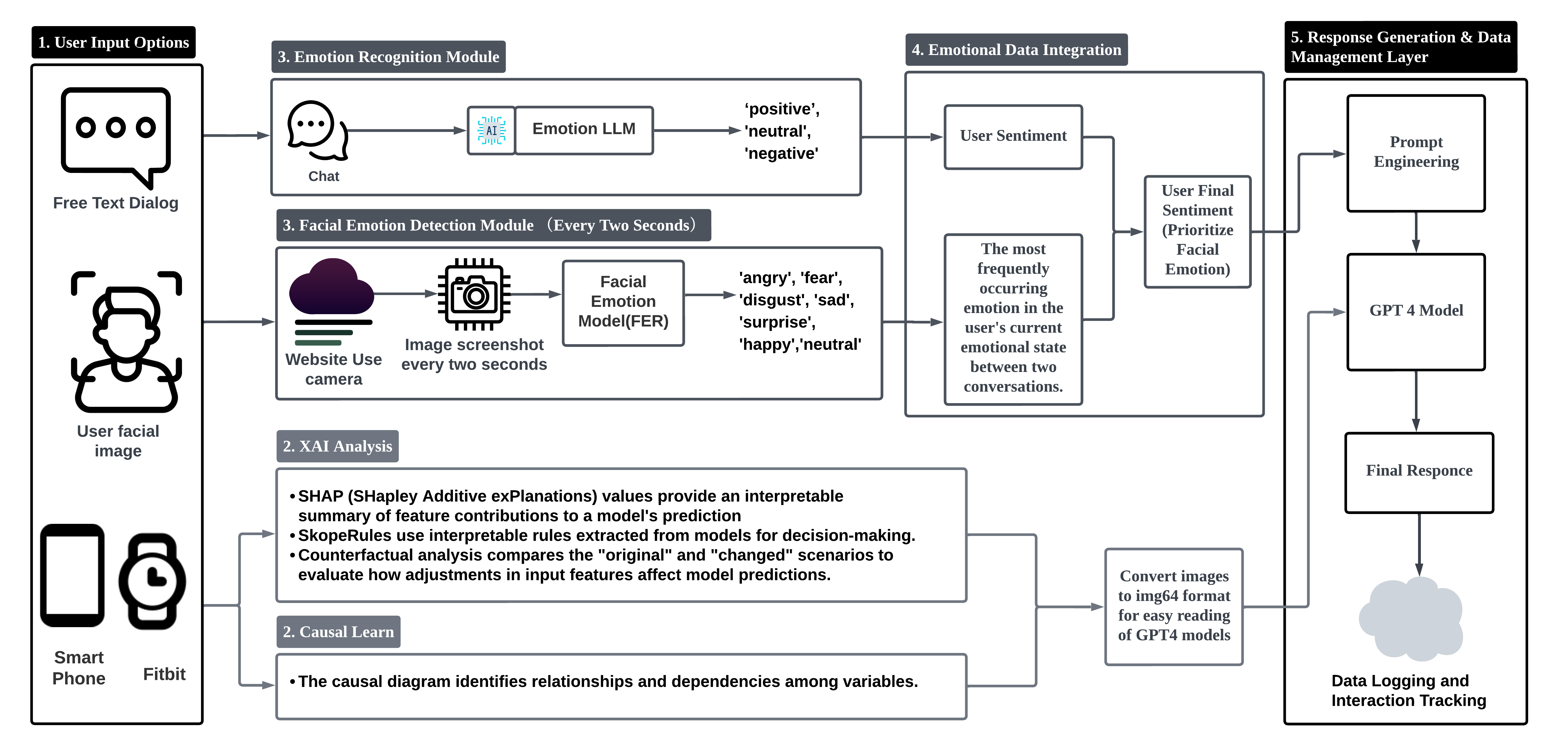}
  \caption{Overview of the Affective Explainable AI-Driven Clinical Decision Support System (AXAI-CDSS)}
  \label{fig: pipeline}
\end{figure*}

In this research, we developed a Web-based interactive application to provide interpretable machine learning analytics, causal-inference and sentiment analysis services. The application allows users (e.g., clinicians) to interact with the system through a natural language interface to access information such as interpretation of model predictions, and results from causal-inference and counterfactual reasoning analyses. These features are integrated into a user-friendly chat interface that makes complex machine learning model decisions understandable to non-technical users. Figure \ref{fig: pipeline} illustrates the system's architecture and data flow, integrating user input, emotion recognition, AI processing, and feedback generation.

\subsection{Data Collection}
The study employed used the AWARE framework app \cite{ferreira2015aware} to collect data from individuals who reported cannabis use with smartphone sensors, such as GPS and accelerometer, \textcolor{black}{alongside Fitbit Charge 2 data for heart rate and step count \citep{chung2020mobile, bae2021mobile, baemohammad}. The data were analyzed to identify patterns of cannabis use, with physiological and behavioral data segmented into 5-minute intervals \cite{bae2025enhancing}.} Key statistics, like heart rate and step count from Fitbit, were extracted to evaluate (i.e., detect) cannabis intoxication. Self-reported cannabis use events were classified into 'intoxicated' and 'not intoxicated' to test the system’s ability to determine and interpret behavioral outcomes from sensor inputs \cite{bae2025enhancing}.  \textcolor{black}{Participants were financially compensated for their involvement, receiving \$75 upon completion of the baseline evaluation and \$25 for participating in the debriefing interview. Furthermore, they earned  \$10 per day if they successfully gathered at least 75\% of the required data, including Fitbit recordings and experience sampling method (ESM) inputs \cite{bae2025enhancing}. Participants who completed the system testing were compensated with a \$20 Amazon gift card.}

For the chat data, user interactions and generated responses were logged in Google Sheets. The data are structured with fields for email address, timestamp, role (user or assistant), and content \cite{kunicki2019keep}. This allows for the persistent storage of conversation history, which can be retrieved and used to maintain context in the ongoing session.

We also conducted a user feedback survey to gather qualitative insights about system performance. The survey evaluated various aspects such as clarity, system usability and personalized recommendations. We asked participants to rate the system based on their own experiences, which helped refine the feedback mechanism and interface design. This feedback was critical in identifying the system's strengths and areas for improvement, especially in terms of user experience and interpretability of AI-driven insights.\textcolor{black}{ Table \ref{tab:participant_data} shows information about the participants.}


\begin{table*}[htbp]
\centering

\begin{tabular}{|c|c|c|c|c|c|c|}
\hline
\textbf{PID} & \textbf{Ethnicity} & \textbf{Gender} & \textbf{Occupation/Area (Years of Experience in the Medical Domain)} & \textbf{Age} & \textbf{Education} \\ \hline
P1 & Asian & M & Pharmaceutical Researcher (10) & 20-29 & Doctoral Degree\\ \hline
P2 & Asian & M & Medical Expert (6) & 20-29 & Master Degree \\ \hline
P3 & Asian & M & Computer Programmer (0) & 20-29 & Bachelor Degree\\ \hline
P4 & Asian & M & Medical Product Manager (10) & 30-39 & Master Degree \\ \hline
P5 & Asian & M & Researcher in Medical Field (6) & 30-39 & PhD Student \\ \hline
P6 & Asian & M & Graduate Student in Psychology (10) & 20-29  & Master Student \\ \hline
P7 & Asian & M & Software Development Engineer (0) & 20-29 & Master’s degree \\ \hline
P8 & Asian & F & Engineer (0) & 20-29 & Master’s degree \\ \hline
P9 & Asian & F & PhD student in Engineering (3) & 20-29 & Ph.D. \\ \hline
P10 & Prefer not to say & M & Business (0) & 30-40 & Master’s degree \\ \hline
P11 & Asian & M & Physician (25) & 50-60 & Postgraduate, MD, PhD \\ \hline

P12 & White & M & PhD student in Engineering (2) & 20-30 & Bachelor’s degree \\ \hline

\end{tabular}

\caption{Participant Demographics and Job Details}
\label{tab:participant_data}
\end{table*}

\subsection{Model Generation}
In this study, we employed a personalized model generation strategy, whereby we constructed a machine learning model that is unique to each cannabis user to ensure that the model accurately captured and reflected individual differences. This approach centers on generating a separate model for each cannabis user, rather than using a common model for all cannabis users' data. Each cannabis user's data is first subjected to initial processing after collection, including operations such as removing duplicate values, filling in missing values (e.g., using the average value as appropriate), and normalizing values to ensure data quality and consistency. In order to eliminate redundancy between features, we also calculated the correlation between features and removed highly correlated features to avoid model overfitting. \textcolor{black}{At the same time, this approach can optimize feature selection, thereby reducing redundancy and improving computational efficiency. Specifically, after correlation analysis, step count statistics (median, maximum, Q1, Q3), sedentary behavior indicators (Q1 sedentary), and battery discharge data (time and power consumption standard deviation, maximum, minimum) were removed. This significantly reduced the complexity of the model without affecting the predictive performance.} For each cannabis user, we extracted the full history of that user from the preprocessed data. This data was used to train a personalized model for each cannabis user \cite{10652070}. For example, in counterfactual analysis, we used decision tree regression models to train each cannabis user individually. This individual data-based training ensures that the model fully captures the user's personalized characteristics and behavioral patterns, rather than being diluted by group characteristics. With this personalized model generation strategy, we not only improve the accuracy of model predictions, but also enhance the user's trust in the model's analytical results. 

\subsection{Interpretability Analysis}
Interpretability analysis aims to provide users of the system (e.g., clinicians) with an understanding of the behavior of the machine learning models. By integrating SHAP (Shapley Additive Explanations) Value Calculation \cite{lundberg2017unified}, Causal Analysis \cite{zheng2024causal}, Counterfactual Analysis \cite{wachter2017counterfactual}, and Rule Extraction \cite{skope_rules}, the system is able to explain the model's decision-making process from multiple perspectives, enhancing the transparency and interpretability of the model.

This system combines a variety of interpretability techniques to help users (e.g., clinicians) understand and scrutinize the behavior and decisions of machine learning models from multiple perspectives. Whether it is global feature importance analysis based on SHAP values, variable interaction impact analysis based on causal inference, hypothetical scenarios generated through counterfactuals, or decision rules extracted through SkopeRules, the system is committed to presenting complex model behaviors to the user in a clear and easy-to-understand manner \cite{10652070}. This multi-level explanatory analysis not only improves the transparency of the model, but also enhances users' trust and understanding of the model results. All XAI-related images are converted to img64 format and transmitted to the GPT-4 model, so that AI can read the information in the image. Figure \ref{fig: dashbord} shows the various forms that XAI takes in the system.


\subsection{Prompt Engineering}

\begin{figure*}
  \centering
  \includegraphics[width=\linewidth]{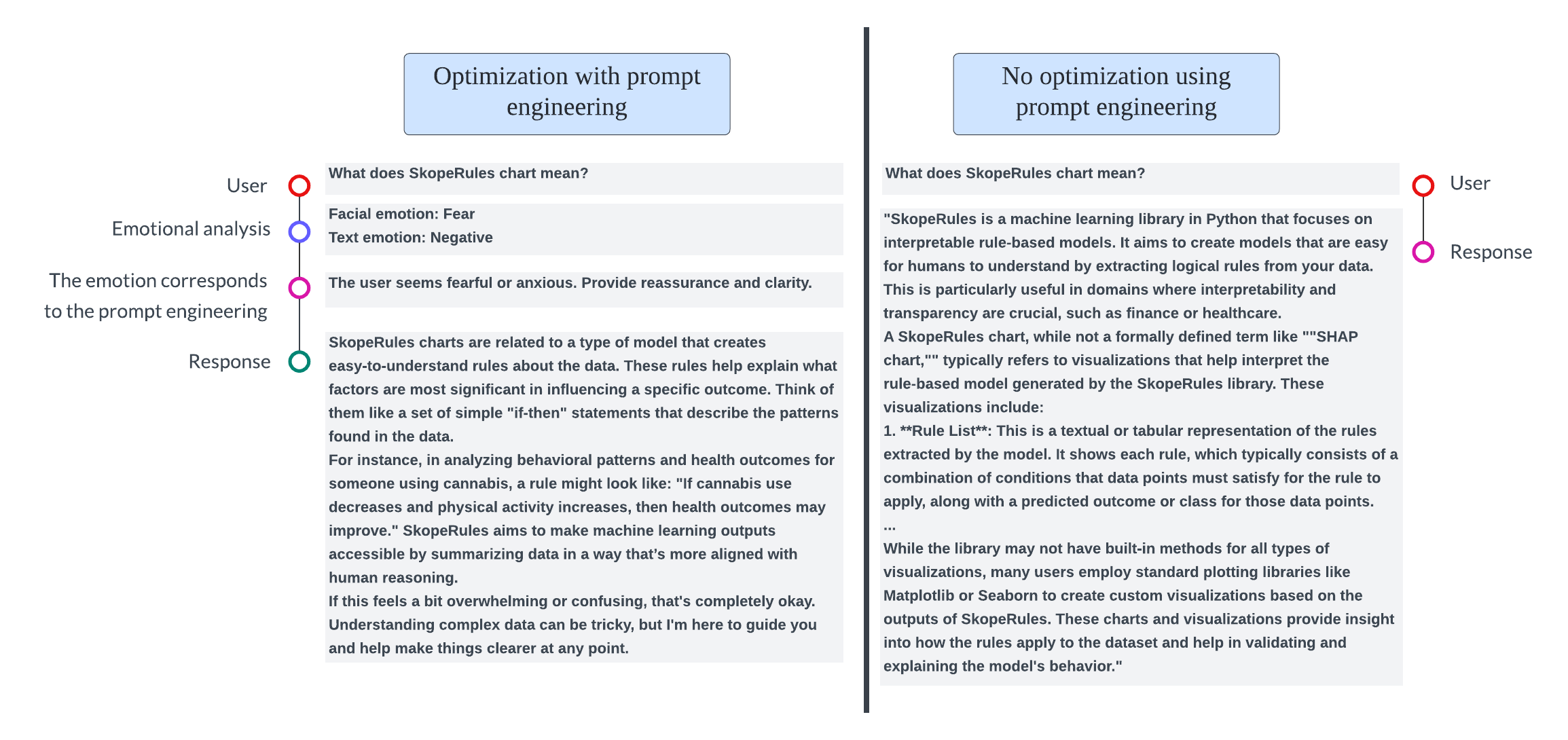}
  \caption{Comparison of Our Proposed Real-Time Emotion-Aware Conversational AI (Left) and Basic LLMs (Right)}
  \label{fig: prompt}
\end{figure*}
In this research, we designed and developed a Web-based interactive dialogue system that can dynamically respond to the emotional state of the clinician using the system. By leveraging prompt engineering \cite{bernardo2023affective}, the system optimizes feedback in dialogue interactions and integrates two complementary components to achieve its core function. The first component uses facial expression recognition technology and text sentiment analysis to identify the emotional state of the user, including anger, disgust, fear, happiness, sadness, surprise, and neutrality. The second component uses text sentiment recognition to generate an emotional analysis of the user's input text. By combining the analysis of the two, we transmit the resulting emotion to the GPT-4 model.

The facial emotion recognition module is a key innovation of the system. It uses a camera-based setup to detect and record the user's facial expressions every two seconds, \textcolor{black}{since the duration of a neutral expression ranges up to 4 seconds \cite{ekman2004emotions}}. The module then identifies the most intense emotional changes between consecutive readings and uses the changes to adjust the system's response. For example, when the system recognizes that the user (e.g., a clinician) is experiencing a negative emotion (e.g., anxiety, confusion, or anger), it adjusts its tone to convey empathy and support, avoiding overly complex terminology while maintaining a respectful and professional tone. Conversely, when the system detects that the user is in a positive emotional state, it responds with an uplifting and affirmative tone, delivered with sincerity and professionalism. For neutral emotional states, the system ensures a professional and clear tone, prioritizing clarity and ease of use. \textcolor{black}{Users do not need to actively chat with the system to extract emotions; instead, the system passively analyzes their facial expressions and short text input in real time. This ensures minimal disruption to their workflow while still allowing the system to adjust its response based on their emotional state. At the same time, the “chat” icon allows users to quickly locate the system's response, thereby reducing confusion when using it.}

\textcolor{black}{In this study, we designed a prompt engineering approach that combines multimodal affective data to ensure that AI interactions are empathetic and context-aware. The system integrates facial expression recognition and text sentiment analysis, prioritizing facial expressions as the primary indicator of user emotions. This is because in many cases, users may suppress emotional cues in text and express true emotions more nonverbally \citep{krystal2019nonverbal,sebe2005multimodal}. However, text sentiment analysis is still used as a secondary function to improve the interpretation of the context. The system assesses the relative change in facial expression rather than the absolute value, to prevent sudden changes in tone due to minor fluctuations in emotion. For example, if a clinician's level of frustration decreases slightly from 100\% to 95\%, they are still frustrated, and the system maintains empathy rather than shifting to neutral. This approach ensures a stable adaptation of the response and avoids overreacting to insignificant changes. We chose GPT-4 primarily because of its unique multimodal capabilities, which enable it to seamlessly handle facial image input and text conversations. Unlike most language models that only process text, GPT-4 is able to analyze images, making it particularly suitable for combining facial emotion recognition with conversational AI. The comparison between our system and a common GPT-4 system is shown in Figure \ref{fig: prompt}.}


\textcolor{black}{In addition to emotionally responsive interactions, the system employs a strategy of simplifying technical details to make XAI explanations more interpretable. Instead of presenting specific numerical values or highly technical expressions}, the system instead uses qualitative and relevant descriptions. For example, instead of reporting the precise quantitative weight for a feature to describe a feature's importance to model prediction, the system conveys feature importance by saying “this feature plays a more important role in the model decision”. Similarly, model output is also expressed in terms of 'high probability' or 'medium impact', rather than precise numerical probabilities. This simplification makes it easier for non-technical users to understand the model's explanations, thereby lowering the cognitive threshold for understanding complex concepts \cite{liao2021human}. The integration of these strategies enables the system to provide more empathetic and contextual interactions, and to improve user engagement. The precise model values are available to the user of the clinical decision support system (CDSS), if requested in a prompt.

\subsection{Dialogue System}

\begin{figure*}
  \centering
  \includegraphics[width=\linewidth]{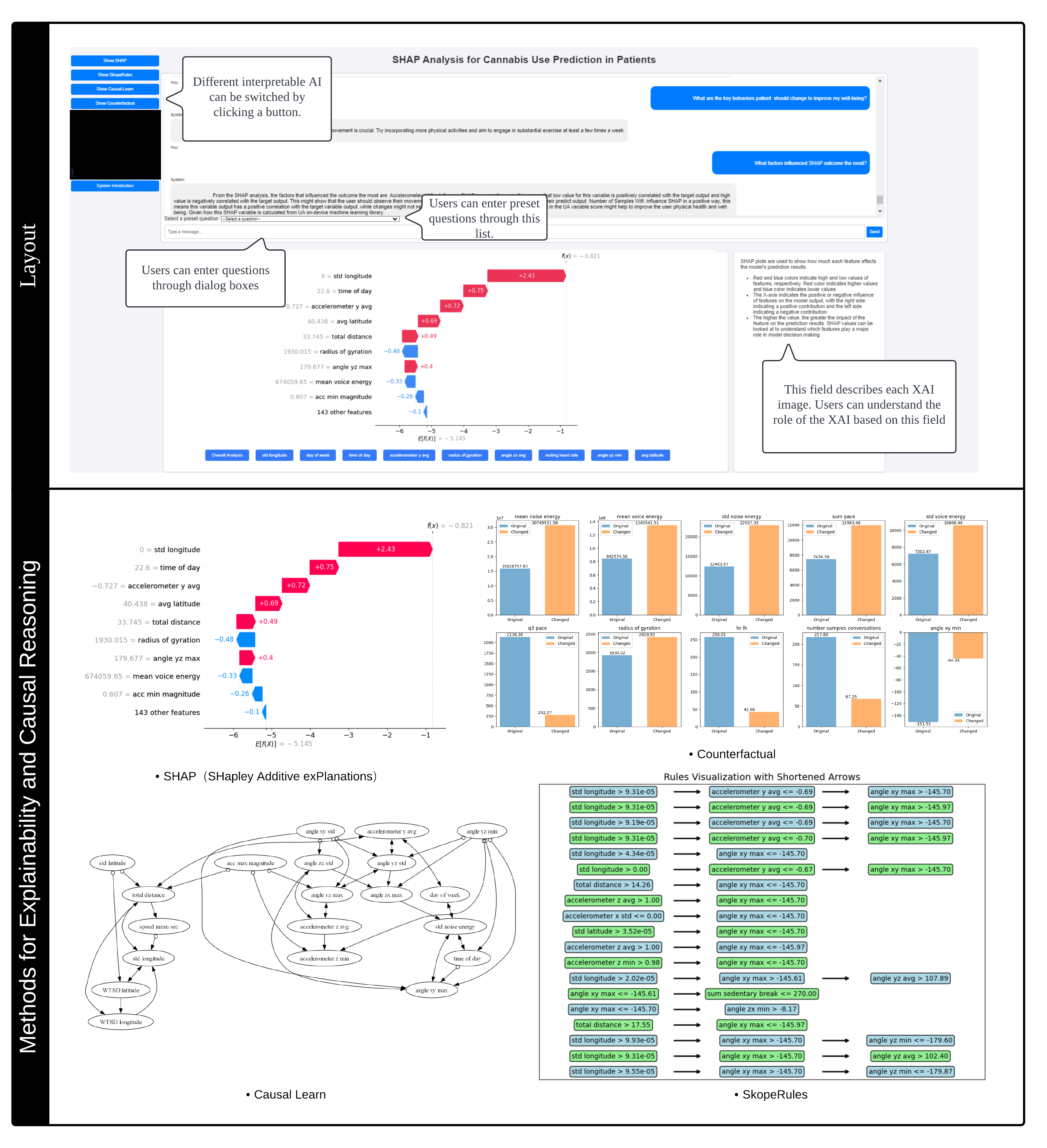}
  \caption{System Demonstration (Top) and Example of Multi-AI Explainability with Causal Reasoning (Bottom)}
  \label{fig: dashbord}
\end{figure*}

 The sentiment detection layer independently processes text sentiment analysis and facial sentiment recognition. Facial sentiment is detected every two seconds, and the system calculates the change in sentiment between user interactions. \textcolor{black}{In our dialogue system, all seven emotions (joy, sadness, anger, neutrality, fear, disgust, surprise) share a percentage distribution, and the one with the highest percentage determines the final classification. Since emotions usually last up to four seconds, but conversations last one to two minutes \cite{ekman2004emotions}, neutral states are not very meaningful for capturing the overall emotional trend of the user. Therefore, neutral emotions are excluded, and the other emotions are ranked according to intensity and frequency. The priority sequence is determined based on the framework of psychological research on emotional arousal \cite{smith1985patterns}, and is arranged in descending order according to the activation intensity of emotional valence. Specifically, negative emotion categories (such as anger and fear) are assigned higher decision-making weights, which is consistent with the priority processing of threats in human emotional response mechanisms. This ensures that the system focuses on the most prominent emotions throughout the interaction. In addition, in order to prevent misclassification, if the user does not show positive or negative emotions, the system will default to a neutral, professional response, rather than force an incorrect emotional label.} The sentiment with the highest frequency is then set as the user's current sentiment. Meanwhile, text input is analyzed separately using a sentiment detection language model to detect sentiment. These results are combined in the response generation layer and submitted to the GPT-4 model to generate empathetic and contextually appropriate responses. The XAI analysis layer integrates data from the cannabis user’s smartphone and wearable devices such as Fitbit to generate explainable insights using SHAP, SkopeRules, and counterfactuals. \textcolor{black}{These insights are presented in a user-friendly visual format within the interface, allowing clinicians to directly view the model results alongside the interaction with the GPT-4 model.} These visual explanations are further transmitted as images to the GPT-4 model, enabling the system to provide detailed, context-aware, and interpretable responses to the clinician user's queries.

The web-based interface enables real-time interaction, where users can either enter custom questions or choose from predefined queries. The system responds to user queries in real time. The system's responses are presented in a natural and empathetic manner, and are dynamically adjusted to the emotional state of the user as detected by a facial emotion recognition module via the camera. This module continuously analyzes emotions in real time, capturing subtle facial cues every two seconds and adjusting the response accordingly. The interface also includes the ability to explore XAI outputs such as SHAP explanations, with the system's user able to toggle between different interpretability methods at the touch of a button. These system outputs are displayed as interactive charts or tables that can be zoomed and repositioned for easy inspection by the user. To make complex insights more understandable, the system avoids technical jargon and instead provides qualitative, user-friendly explanations, bridging the gap between technical insights and clinical practice decisions. By combining multimodal sentiment recognition with explainable AI, the system not only enhances trust and usability, but also provides personalized explanations based on clinicians' needs and context. Figure \ref{fig: dashbord} shows the interface of our system.

\section{Result}


\textcolor{black}{Our analysis utilized multiple XAI techniques to reveal patterns of behavior associated with cannabis use, highlighting key differences in activity levels, environmental conditions, and physiological responses. Notably, some participants exhibited frequent motion fluctuations, indicating an increase in their level of agitation during self-reported cannabis use. This observation was in line with changes in accelerometer readings, where elevated motion changes were detected, particularly in the dominant axis. Noise level measurements indicated that a subset of users were frequently exposed to high background noise, further substantiating the likelihood of cannabis use in social or public settings. This model uses a dataset containing a large number of negative samples during both the training and testing stages to improve the stability of its performance in scenarios with imbalanced category distributions. The dataset more realistically simulates the imbalanced distribution characteristics of categories in real-world scenarios by making the number of negative samples (samples that do not use cannabis) significantly greater than that of positive samples. }

\textcolor{black}{The cannabis use prediction in this project specifically refers to the identification (i.e., detection) of acute cannabis intoxication events based on physiological and behavioral sensor data. Data were collected from 34 young adults (18–25 years old) recruited from the community who reported current cannabis use. Data included heart rate, step count, and activity information from a Fitbit, as well as contextual data such as GPS and phone usage from the AWARE mobile sensing app. The system's model-prediction framework involved training a separate machine learning model for each participant using their sensor data (smartphone and Fitbit). Algorithms such as XGBoost and decision trees and rule-based models such as SkopeRules were tested, and the best-performing model for each individual was selected based on metrics such as accuracy, precision, recall and F1 score. The average accuracy for all participants exceeded 0.85, demonstrating strong predictive performance.} 

These explanatory AI tools can help us understand how models make predictions in light of specific features. One of our main challenges in this process was to translate these complex model outputs into feedback that, for example, users of the system, such as clinicians, could understand and apply. During the evaluation process, we not only focused on the predictive accuracy of the model, but also delved into the impact of prompt design on model interpretability and system user experience. By comparing the results with and without the use of prompt design, we found that prompt design improves system performance significantly. Without hints (i.e., effective prompts), the system feedback is often general and lacks specificity, making it difficult for the system's users to quickly understand the meaning of complex model results. However, with effective prompt design, we were able to break down complex model logic into more understandable parts, thus providing more detailed, precise, and understandable feedback to the system's users.


\begin{table*}[htbp]
\centering
\caption{Survey Questions}
\label{tab:survey_questions}
\scalebox{0.9}{
\begin{tabular}{|p{0.2\textwidth}|p{0.75\textwidth}|}
\hline
\textbf{Section} & \textbf{Question} \\
\hline
\multirow{4}{=}{System Usability \cite{brooke1996sus}} & Q1: How easy was it to interact with the system using natural language? (1 - Very Difficult, 10 - Very Easy) \\
& Q2: How intuitive was the user interface for navigating through your data and the system's insights? (1 - Not Intuitive, 10 - Very Intuitive)  \\
& Q3: How satisfied were you with the system’s response time in delivering insights? (1 - Very Dissatisfied, 10 - Very Satisfied) \\
& Q4: Does the system's answer correctly interpret cannabis user data? (1 - Very Dissatisfied, 10 - Very Satisfied)\\
\hline
\multirow{3}{=}{Personalization and Relevance  \cite{cornellux}} & Q1: How do system insights and recommendations correlate with improved personal health and wellness for patients? (1 - Not Relevant, 10 - Very Relevant)\\
& Q2: To what extent did you feel the insights were personalized to cannabis patient's individual data and behaviors? (1 - Not Personalized, 10 - Highly Personalized) \\
& Q3: Did the system adequately consider patient's current condition when providing insights (e.g., heart rate, sleep patterns)? (1 - Not at All, 10 - Completely)\\
\hline
\multirow{5}{=}{Clarity and Comprehensibility \cite{borsci2022chatbot}} & Q1: How clear and understandable were the explanations provided by the system (e.g., SHAP values, causal diagrams)? (1 - Very Confusing, 10 - Very Clear)\\
& Q2: How helpful were the system’s visualizations (e.g., SHAP plots, causal diagrams) in aiding your understanding of the results? (1 - Not Helpful, 10 - Very Helpful)\\
& Q3: Did you find the response generated by the system easy to follow? (1 - Very Difficult, 10 - Very Easy) \\
& Q4: Does this response contain evidence of correct personalization, reference appropriate user data, or correctly refuse to answer when such data is missing? (1 - Very Difficult, 10 - Very Easy)\\
& Q5: Does the system's response demonstrate correct personalization, reference relevant user data appropriately, and correctly refuse to answer when the required data is missing? (1 - Very Difficult, 10 - Very Easy) \\
\hline
\multirow{4}{=}{System Benefits \cite{el2023factors}} & Q1: How beneficial do you believe the insights were in improving your understanding of patient health behaviors (e.g., sleep, exercise)? (1 - Not Beneficial, 10 - Very Beneficial)\\
& Q2: To what extent do you believe the system's recommendations will help you improve patient health outcomes (e.g., better sleep, increased physical activity)? (1 - Not Likely to Help, 10 - Very Likely to Help) \\
& Q3: Would you recommend this system to others looking for patient health insights from wearable data? (1 - Definitely Not, 10 - Definitely Yes) \\
& Q4: Did the system's response avoid misleading you? (1 - Definitely Not, 10 - Definitely Yes)\\
\hline
\multirow{2}{=}{Overall Satisfaction \cite{el2023factors}} & Q1: How satisfied are you with the overall quality of the insights and recommendations provided by the system? (1 - Very Dissatisfied, 10 - Very Satisfied)\\
& Q2: How likely are you to continue using this system to monitor and improve patient health in the future? (1 - Not Likely, 10 - Very Likely)\\

\hline

\end{tabular}
}
\end{table*}

This study evaluates the performance of two systems. The first is the optimization system introduced in this paper, which incorporates both facial emotion recognition and text-based sentiment analysis to provide personalized and emotionally intelligent responses. The second is not an optimization system, created for the purpose of comparison by removing the facial emotion recognition and text-based sentiment analysis components. Both systems share the same interface layout to ensure consistency, with the comparison focusing on usability, personalization, and clarity through user evaluations conducted under controlled conditions. \textcolor{black}{The two system interfaces are shown in Figure \ref{fig: dashbord}.} To evaluate both systems, a comprehensive survey was conducted, see Table \ref{tab:survey_questions}, which included five key dimensions: system usability, personalization and relevance of insights, clarity and understandability, system strengths, and overall system user satisfaction. The survey questions were designed to measure the ease of use of interactions with the system, the appropriateness of personalized feedback, and the clarity of system-generated insights.

During participant recruitment, we posted recruitment announcements on online platforms and screened eligible participants, including medical practitioners, health care providers, and system users with certain needs for emotion recognition. All participants signed informed consent forms and were informed of the data usage and privacy protection measures. The participants’ backgrounds covered related fields such as clinical decision-making and healthcare to ensure broad applicability of the experimental results. Participants logged in to the system and accessed the designated interface to complete interactive tasks with the system (e.g., inputting health-related questions in natural language). After each interaction, the system performed emotion recognition and generated interpretative feedback. Experimental data are automatically recorded via a predefined Google Sheets API, including participant input, system feedback, and sentiment classification results. Participants were also asked to rate various aspects of the system using a questionnaire after each test.

\begin{figure*}[htbp]
    \centering
    \begin{subfigure}[b]{0.45\textwidth} 
        \centering
        \includegraphics[width=\textwidth]{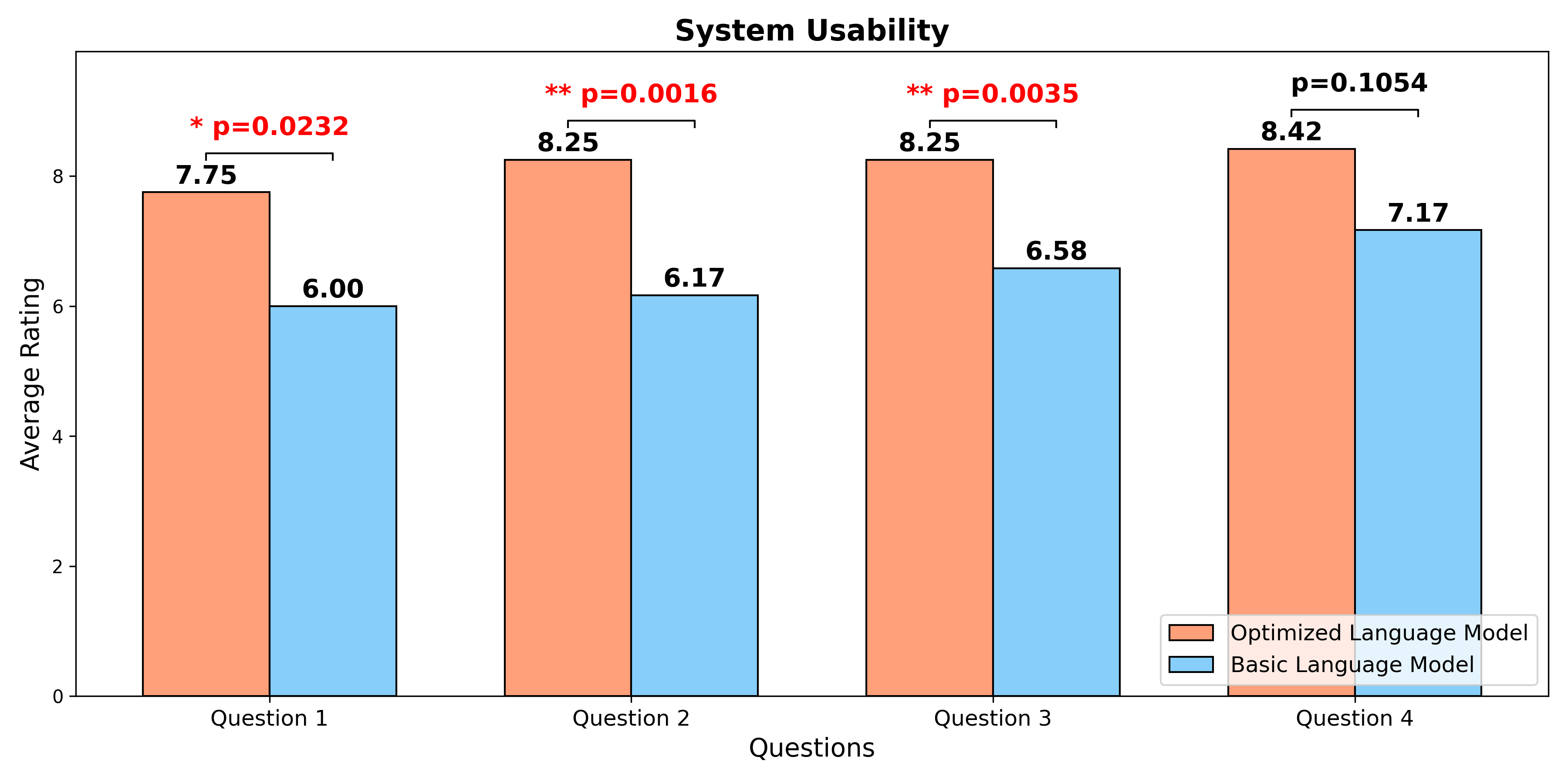}
        \caption{System Usability}
        \label{fig:usability}
    \end{subfigure}
    \hfill
    \begin{subfigure}[b]{0.45\textwidth} 
        \centering
        \includegraphics[width=\textwidth]{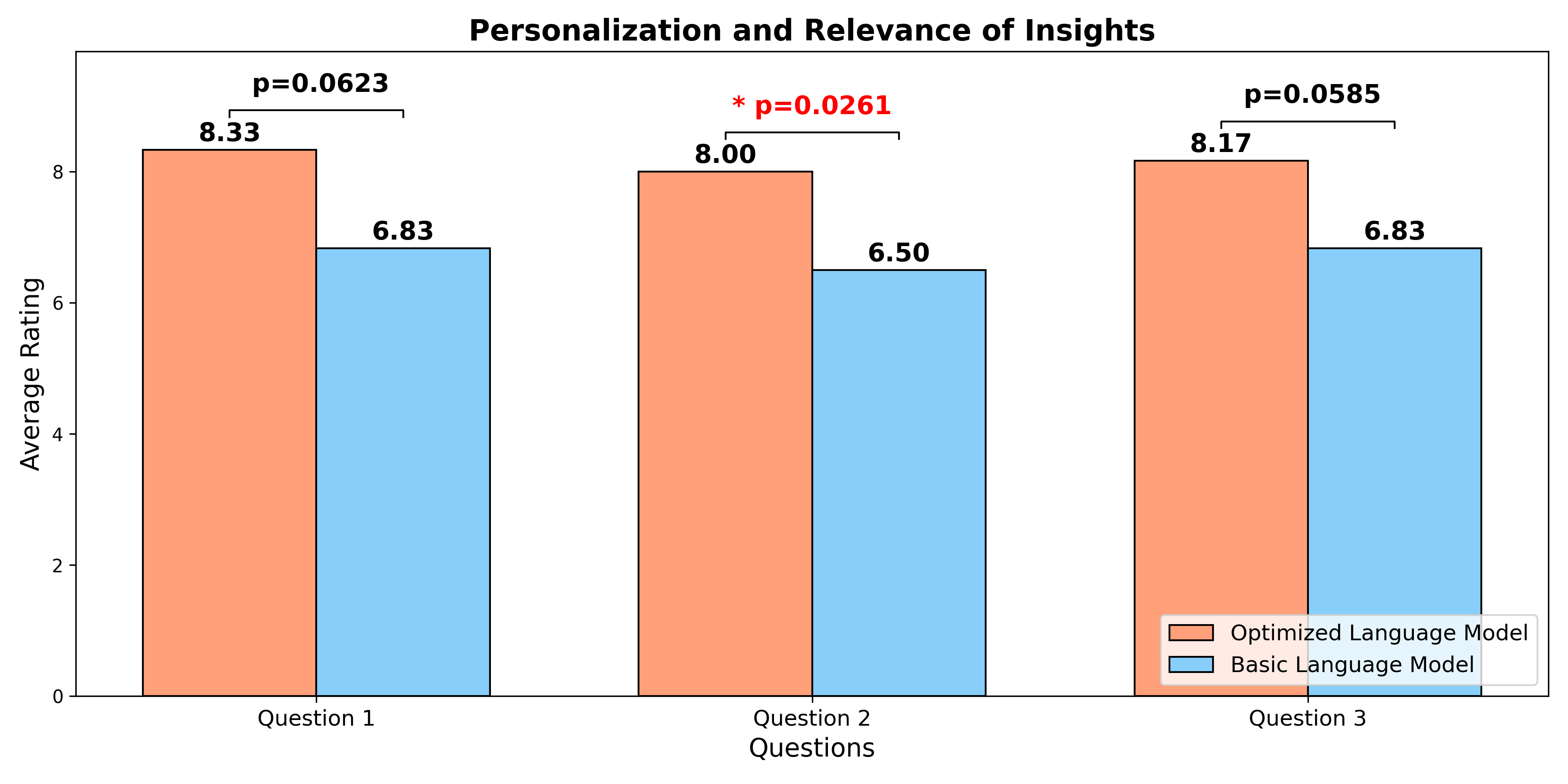}
        \caption{Personalization and Relevance}
        \label{fig:personalization}
    \end{subfigure}
    
    \vskip\baselineskip 

    \begin{subfigure}[b]{0.45\textwidth} 
        \centering
        \includegraphics[width=\textwidth]{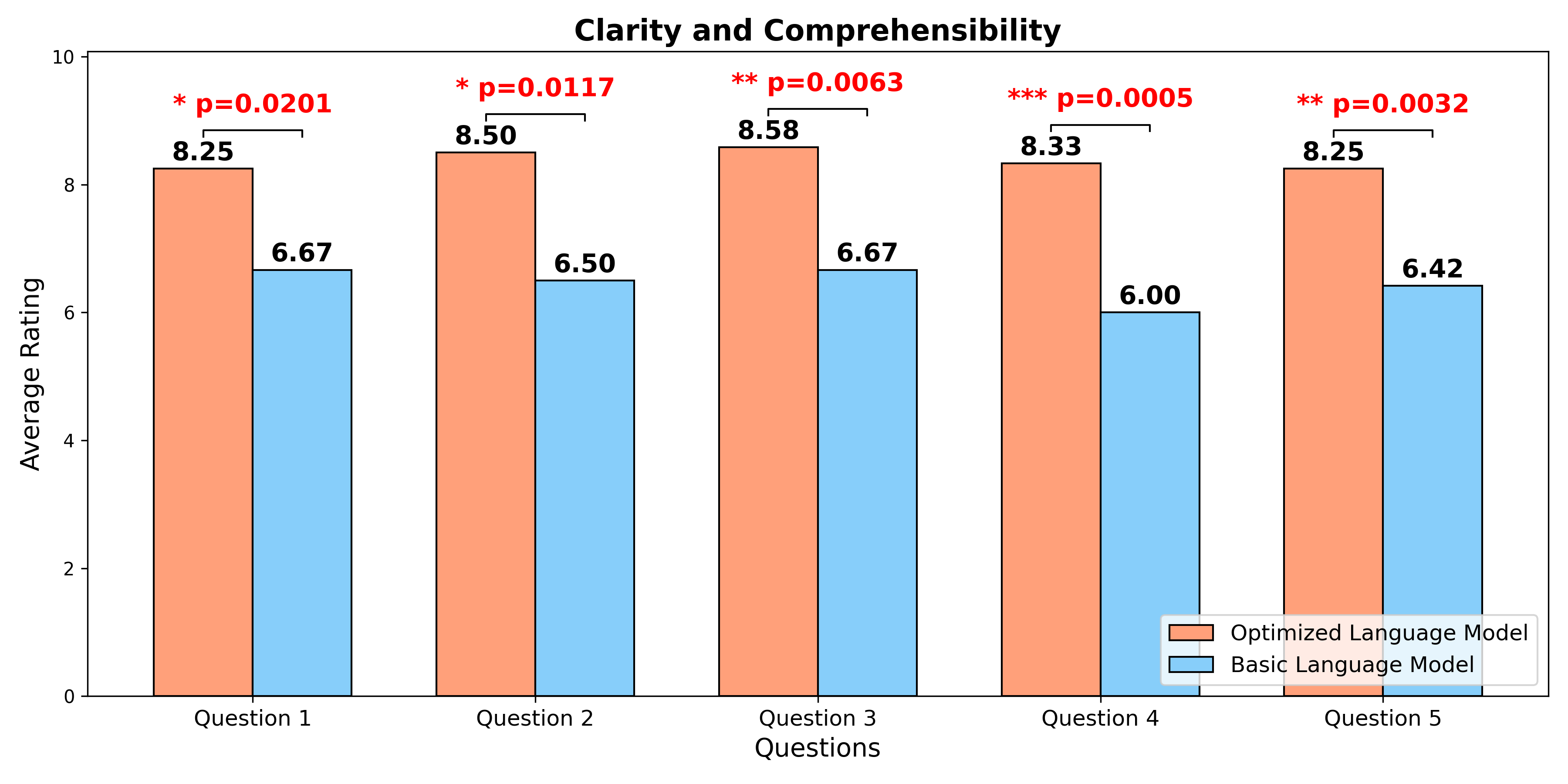}
        \caption{Clarity and Comprehensibility}
        \label{fig:clarity}
    \end{subfigure}
    \hfill
    \begin{subfigure}[b]{0.45\textwidth} 
        \centering
        \includegraphics[width=\textwidth]{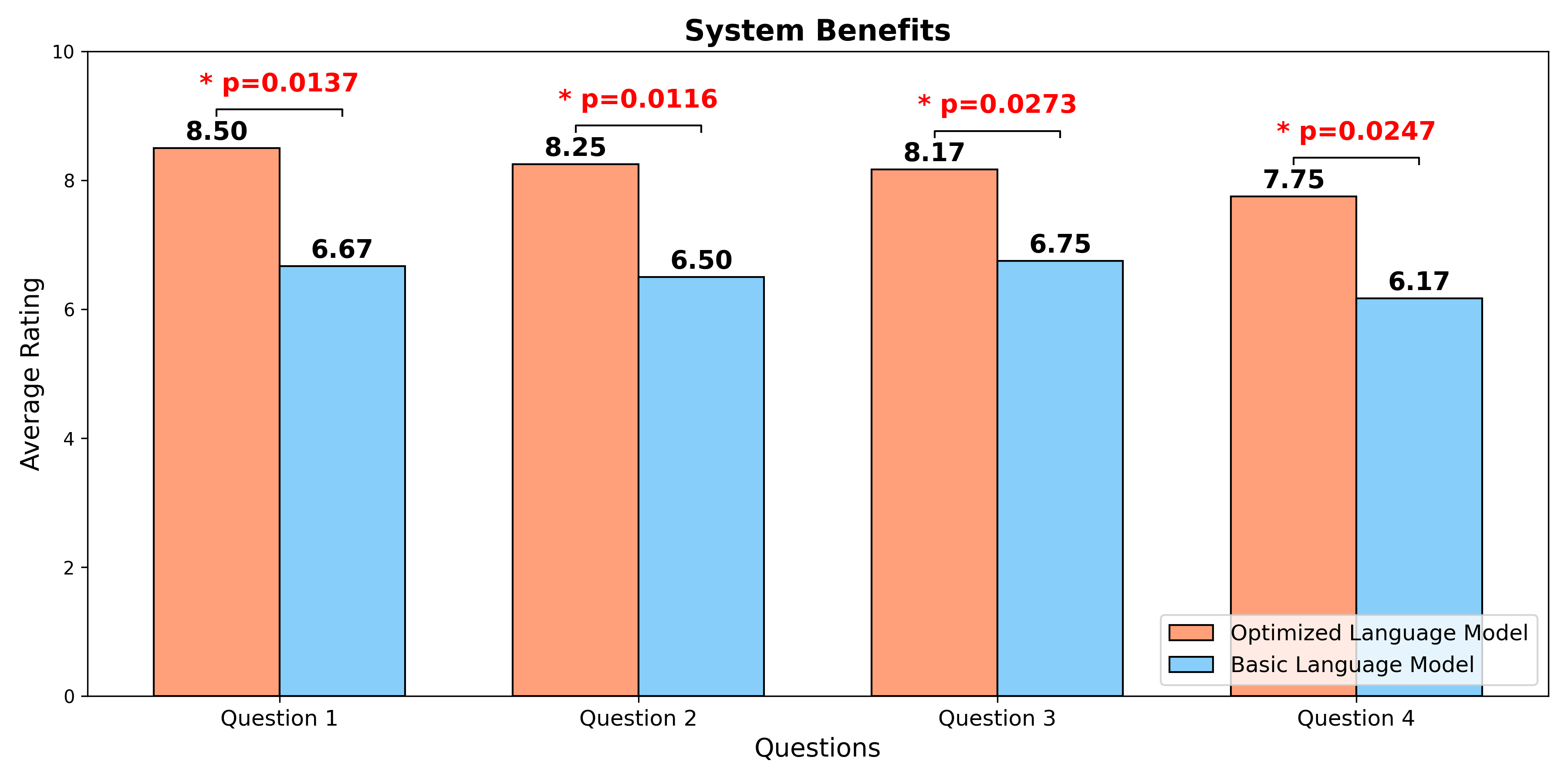}
        \caption{System Benefits}
        \label{fig:benefits}
    \end{subfigure}
    
    \vskip\baselineskip 

    \begin{subfigure}[b]{0.45\textwidth} 
        \centering
        \includegraphics[width=\textwidth]{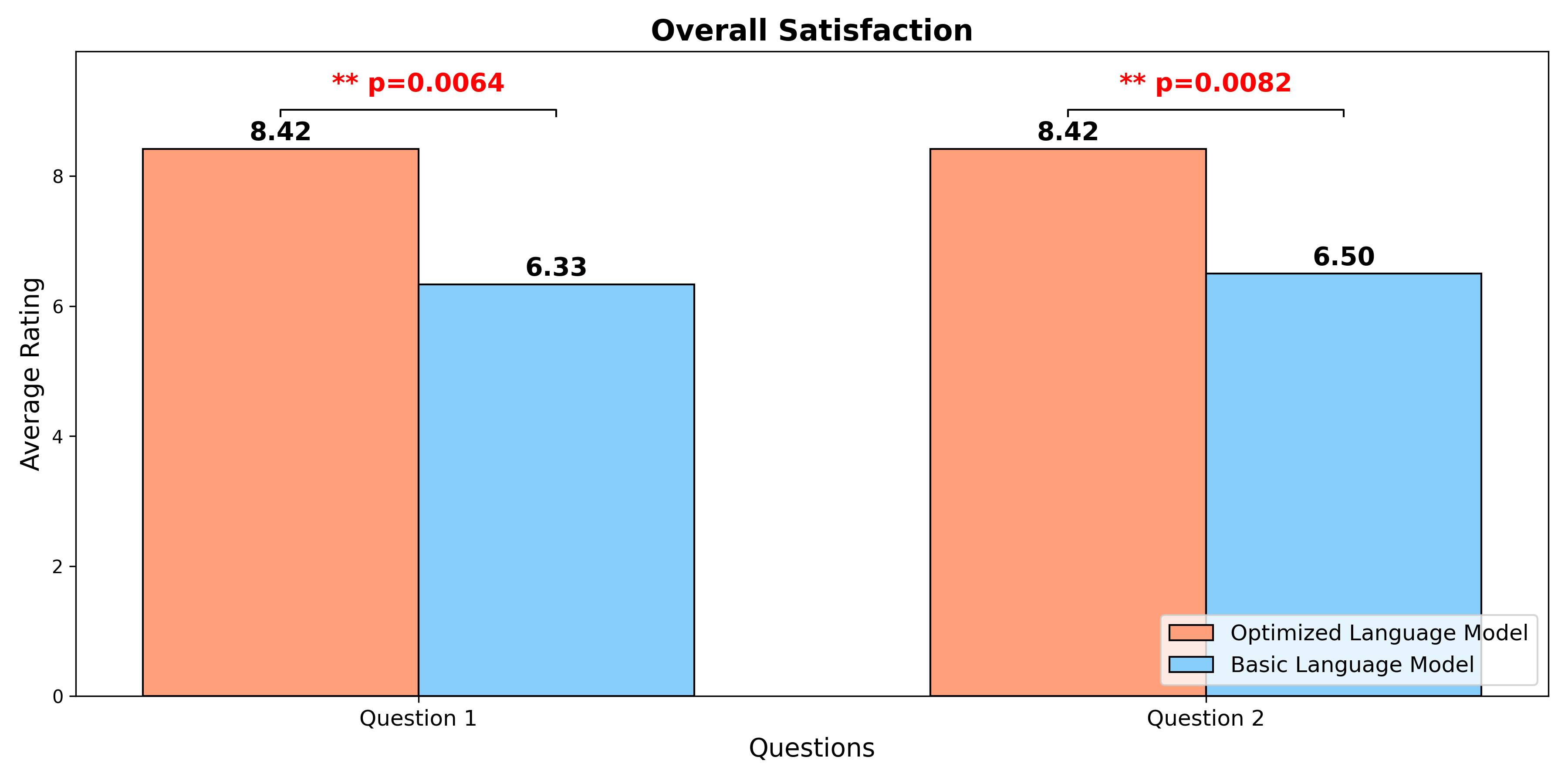}
        \caption{Overall Satisfaction}
        \label{fig:satisfaction}
    \end{subfigure}

    \caption{Comparative Results across Different Metrics}
    \label{fig:comparison}
\end{figure*}

In addition, the comparative results of the systems on these dimensions are illustrated in Figure \ref{fig:comparison} to show how the optimized system significantly outperforms the unoptimized model in each of the dimensions of clarity, personalized feedback, and overall system usability. The light orange colour in Figure \ref{fig:comparison} represents our system language model and the light blue colour represents the base language model. Each bar corresponds to a specific question (e.g., ‘Question 1’, ‘Question 2’), and the connecting line at the top of the bar and the labelled p-value indicate the significant difference between the two sets of scores. Significance markers (e.g. *, **, ***) are coloured in red to indicate the significance levels $p < 0.05$, $p < 0.01$ and $p < 0.001$, respectively. p-values are marked in black if they are not statistically significant. This figure allows visualising the difference in performance between the two models on different problems and their significance. These evaluations help us better understand the role of prompt engineering and sentiment analysis models in improving the interpretability of the system, as evidenced by clearer feedback and more relevant suggestions, as well as improving the overall user experience through more intuitive interactions and faster response times.

In terms of system usability (Figure \ref{fig:usability}), the role of prompt engineering is crucial in enabling the system to more accurately understand the user's natural language input and provide intelligent feedback accordingly. Sentiment analysis modeling also plays a role in this dimension by enabling the system to adjust the interaction based on the emotional context of the user's input, ensuring that the system's feedback is not only accurate but also emotionally connected to the user.

The optimization system's strengths in personalized recommendations were more effective than the base system. \textcolor{black}{ On Question 2 in Figure \ref{fig:personalization}, a paired t-test was performed to examine the difference between subjective reports for the optimized and basic systems. The analysis revealed a significant improvement, t(11) = 2.57, $p$ = 0.026, with a medium effect size (Cohen's d = 0.74), suggesting a meaningful intervention effect. The optimized system scored 8.00 (SD = 1.35), compared to the basic system's 6.50 (SD = 1.93). These scores were based on a 10-point scale, where 1 represented "not relevant at all" and 10 represented "highly relevant".} This suggests that the optimization system is better able to provide personalized recommendations with higher relevance based on the user's specific health condition. This is largely due to prompt engineering, which allows the system to be more flexible and more tailored to the participant's mood in adjusting the feedback content and generating analytics that are more relevant to the cannabis use.


For clarity and comprehensibility,  Figure \ref{fig:clarity} shows that the optimized language modeling system using prompt engineering and sentiment analysis outperformed the base language modeling system on all questions. The optimized system achieved a score of 8.58 out of 10 on question 3, "Clarity of System Explanation," compared to 6.67 out of 10 for the base system. Statistical analysis revealed a substantial improvement, t(11) = 3.36, $p$ = 0.063, with a large effect size (Cohen's d = 0.97). This difference indicates that the optimized system utilizes prompt engineering techniques to simplify the output of complex explanatory AI into easy-to-understand text. This allows users to more intuitively understand the causes and logic of the model's predictions, avoiding confusion caused by technical jargon or complex analysis results.

The system benefits dimension (Figure \ref{fig:benefits}) evaluates whether the system had a positive impact on the cannabis user's health behaviors and decisions. \textcolor{black}{Question 2 highlights how the optimization system enhances the clinician's capacity to improve patient health outcomes, such as better sleep or increased physical activity, through personalized recommendations. This result demonstrates the potential of the optimization system to support clinicians in promoting better health outcomes for individuals who report cannabis use, even without direct interaction between the system and the cannabis user.}

In Figure \ref{fig:satisfaction} on overall satisfaction, the results indicated a notable improvement, t(11) = 3.35, $p$ = 0.006, with a strong effect size (Cohen's d = 0.97), highlighting the enhanced impact of the optimized system on user satisfaction.. The optimization system using prompt engineering is better able to provide system users with personalized feedback and specific guidance, allowing them to have a higher level of trust and acceptance of the system-generated suggestions.

The optimization system consistently scored highly in terms of clarity and understandability, which indicates that the integration of a bilingual model and sentiment analysis effectively supported users' understanding of the insights generated by the AI. This clarity is further validated by overall user satisfaction, which indicates that most users are satisfied with the insights and recommendations provided by the system. \textcolor{black}{Meanwhile, the integration of real-time sentiment recognition enables the system to dynamically adjust its response, thereby enhancing its personalization capabilities. The system is capable of detecting a range of emotions, including 'angry', 'fear', 'disgust', 'sad', 'surprise', 'happy', and 'neutral'. This wide range of detectable emotions ensures that the system remains useful and adaptive even in low-stress contexts, effectively utilizing emotion analysis to personalize interactions. Personalization and Relevance are perceived as an important advantage, with a median personalization score of 9 in terms of relevance to the individual cannabis user's health and wellness needs, specifically reflecting the system's ability to provide personalized feedback based on the clinician's prompts. However, the variability in personalization scores indicates differences in system user experience, which suggests challenges in maintaining consistency of case-specific insights.} This variability in personalization scores highlights the need for better judgment mechanisms to further tailor responses based on real-time sentiment data. Sentiment-driven responses help to facilitate interactions between clinicians and AI systems, as evidenced by high usability scores, with median ratings for ease of use and intuitiveness between 8 and 9. The system's ability to accurately interpret reference data and adapt response tone based on user sentiment leads to a positive, empathetic user experience that helps to increase engagement and user satisfaction.
\section{Discussion} 

This study introduces a novel approach to integrating multiple models in a clinical decision support system (CDSS) customized for managing cannabis use disorder (CUD) \citep{jorke2024supporting,rajashekar2024human}. The combination of sentiment analysis with interpretable AI explanations generates an emotion adaptive and more transparent CDSS to help guide evidence-based decision making. Traditionally, many healthcare providers may be skeptical of AI systems, and reluctant to rely on AI to support critical clinical decisions \cite{mennis2023cannabis}. By employing interpretable models such as SHAP and rule-based systems, the system provides clinicians with a clear and more understandable view of how specific clinical decisions are made based on multimodal patient data (e.g., self-report, sensor), thereby increasing the probability of trust and adoption of system recommendations \citep{thapa2023chatgpt,park2024assessing}.

Also, in addition to improving decision transparency, the system incorporates sentiment analysis that adjusts system responses based on a healthcare provider's emotional state when using the system \cite{chhetri2023machine}. In busy, high-stress environments, a clinician's immediate emotional state could negatively impact their ability to make sound decisions \cite{kozlowski2017role}. By recognizing and responding to these emotional cues in real time, a decision support system could provide more empathetic feedback to the clinician, thereby reducing cognitive load and potentially preventing burn out. \textcolor{black}{Our findings suggest that effective prompt engineering improves the interpretability of AI by guiding responses towards structured, relevant, and context-aware output. Although the multimodal capabilities of GPT-4 were used in this study, the principles of effective prompt engineering can be generalized to other language models. Models such as Gemini, Claude, and LLaVA could also benefit from structured input, although their performance varies in medical reasoning and multimodal integration. Future work will include evaluating these models using the same prompt design framework to assess their suitability and effectiveness in clinical XAI applications.}

In high-pressure healthcare environments, health care providers often experience a range of emotions that can affect their decision-making processes. The addition of facial emotion recognition enables the system to detect signs of fatigue, stress or frustration that may not be apparent from just text exchanges with the system. For example, a clinician interacting with the system who exhibits subtle expressions of anxiety may receive more calming and reassuring feedback with the inclusion of facial expression recognition \cite{ghandeharioun2019emma}. This emotionally adaptive interaction not only helps reduce cognitive burden, but also could help alleviate burnout, a key issue in healthcare settings \cite{palfi2024adaptive}. The CDSS's enhanced emotional intelligence promotes a more empathetic human-machine interaction, which is critical for building trust and reliance in AI-assisted tools. This empathetic interaction can increase system adoption rates as clinicians feel more understood and supported in their clinical work.

In addition, facial expression recognition addresses some of the limitations of previous iterations of the system. While text sentiment analysis can provide valuable insights, it may fail to capture the full complexity of human emotions, particularly in environments where verbal communication is limited or stressful enough to inhibit the expression of emotive language. Facial expressions often convey nuanced emotional information, which can result in possible misclassification of emotions. If accurately detected and interpreted, emotion recognition can significantly improve the responsiveness of the system. However, facial expression recognition also presents challenges, particularly in terms of data privacy and security. The system processes sensitive biometric data, raising ethical questions about consent and the protection of personal information. Ensuring the confidentiality and integrity of facial data is critical. Future work should focus on implementing encryption techniques and exploring approaches such as federated learning, which processes data on local devices rather than transmitting it to a central server. This approach reduces the risk of data breaches and unauthorized access.

The addition of facial emotion recognition and text sentiment recognition significantly enriches the CDSS, enabling it to provide more empathetic and context-sensitive feedback. This advancement improves the user experience for clinicians, could support more effective evidence-based decision-making, and strengthen the trusting relationship between healthcare providers and AI systems. By addressing the challenges of increasing accuracy of recommendations, data privacy and ensuring diversity and representativeness in the training data, the system can further improve its adaptability and effectiveness in a variety of clinical settings.

\textcolor{black}{One limitation of our system is that it requires a stable internet connection to use GPT-4 for real-time processing. In clinical settings, an unstable network may affect response time and overall usability. To address this issue, future work will explore integrating lightweight local language models to provide offline backup functionality to ensure that the system remains operational even in low connectivity environments. In addition, while our system currently prioritizes facial expression recognition, further research is needed to evaluate the impact of integrating other context-aware features to improve the accuracy of emotion extraction. Future iterations will also investigate other multimodal models to balance the performance and computational efficiency of clinical decision support.}

\textcolor{black}{At the same time, training a unique model for each user requires a significant amount of computation, which motivates us to explore other methods, such as embedding-based modeling. Future work will further investigate hybrid modeling strategies, such as transfer learning and adaptive fine-tuning, to optimize computational requirements and prediction accuracy. In addition, comparative analyses with other personalization techniques, including clustering-based methods or meta-learning, may provide further insights into improving the adaptability of models across users.}

\section{Conclusion}

This research proposes an innovative approach in a clinical decision support system (CDSS) tailored to manage cannabis use disorder (CUD). By combining facial emotion recognition and text sentiment analysis with explainable AI-driven explanations, the system addresses long-standing challenges of AI in healthcare applications. These challenges include the opaque "black box" nature of many AI models and the historical lack of emotional intelligence in AI interactions with clinicians. The addition of facial emotion recognition capabilities enhances the system's ability to detect and respond to the emotional state of healthcare workers in real time. This affective intelligence design is particularly important in clinical settings. By providing supportive and emotionally-appropriate feedback to health care providers using the system, the system could reduce stress and cognitive load \cite{kozlowski2017role}, thereby improving the overall user experience and fostering trust between clinicians and AI tools. 

Furthermore, the use of interpretable AI models, such as SHAP and rule-based systems, can improve the transparency and comprehensibility of the system's recommendations. This transparency is essential for increasing clinicians' trust in and reliance on AI assistance. By revealing the decision-making process behind a particular decision, the system can help reduce concerns of healthcare professionals regarding the "black box" of AI-based recommendations and promote its adoption. By addressing the technical issue of decision transparency and the emotional dimension of clinician support, the system developed in this study shows promise for improving cannabis use disorder management.
\section{Acknowledgment}
This study was supported by the National Institute On Drug Abuse of the National Institutes of Health under Award Numbers U01DA056472 and R21 DA043181.

\bibliographystyle{IEEEtran}
\bibliography{software}

\end{document}